**Title:** A New Method for Partial Correction of Residual Confounding in Time-Series and other Observational Studies

**Authors:**    W. Dana Flanders, Matthew J. Strickland and Mitchel Klein

**Acknowledgement Section**

 "This publication was supported by US EPA grant R834799.  This publication's contents are solely the responsibility of the grantee and do not necessarily represent the official views of the US EPA. Further, US EPA does not endorse the purchase of any commercial products or services mentioned in the publication."

Author Affiliations: Emory University, Rollins School of Public Health, Department of Epidemiology, 1518 Clifton Road, Atlanta, GA 30322 (W. Dana Flanders and Mitchel Klein); Emory University, Rollins School of Public Health, Department of Biostatistics and Bioinformatics, 1518 Clifton Road, Atlanta, GA 30322 (W. Dana Flanders); Emory University, Rollins School of Public Health, Department of Environmental and Occupational Health, 1518 Clifton Road, Atlanta, GA 30322 (Matthew Strickland , Mitchel Klein)

**Introduction**

Confounding is one of the main threats to validity of observational studies. It is a mixing of the effects of an extraneous factor with those of the exposure of interest so as to distort observed associations(1, 2). It's expected if certain causal patterns (confounding paths) are present in a causal graph that reflects the causal relationships (2-4).  Confounding is controlled analytically by adequate stratification or modeling covariate effects so as to block the confounding path. Residual confounding is confounding that remains even after attempts to control it. It can be present if confounders are unmeasured, miss-measured or their form is miss-specified (2, 5). Thus, sensitivity analyses to assess the potential impact of residual confounding can be important (2, 6, 7). In time-series studies of the short-term health effects of environmental exposures, confounding is often controlled, as least partly, by including in the model covariates such as day-of-week, temperature and humidity, and by including splines or other terms for time that can control for unmeasured factors that co-vary smoothly over time.

By considering temporality and causal relationships, we previously showed (8-10) how to test for residual confounding or other model-miss-specification. The test involves adding to the final model an indicator variable with two basic properties: First, the indicator should be independent of disease in a correctly specified model; in particular, it should neither cause nor be caused by the disease. Second, it should be associated with the exposure of interest and like the exposure, with unmeasured confounders. Plausibility of these properties can be evaluated by considering causal diagrams that summarize tenable causal relationships.

The test for confounding is conducted by adding an "indicator" variable to the model to be evaluated. We refer to the final model, but with the indicator added as the "extended" model. An association between the indicator and disease in the extended model suggests model miss-specification. Ability of the indicator to detect confounding was justified using causal considerations, much like those for negative exposure controls (11) and Granger causality (12). In environmental time-series studies, we argued that the exposure level after the event has already occurred is a candidate for such an indicator because of the requirement that a cause precedes the disease, and because health events do not, at least in the short term, affect ambient levels of many environmental pollutants, like air pollution. Thus, future levels of the exposure cannot cause disease and should not be affected by it partially justifying one of the key properties needed for the indicator (8, 10).

To detect residual confounding, Lipsitch et al (11) proposed use of a negative exposure control, similar to but technically slightly different from the indicator described above. They also described a method to correct for residual confounding in sensitivity analyses when stratification, but not a regression model, is used to control for known confounders. Tchetgen (13) also suggested a method to correct for confounding using negative outcome controls, although it potentially requires assumption of a deterministic relationship between the dependent variable and its causes. Nevertheless, the problem of how to correct for residual confounding in regression analyses by using an exposure-based confounding indicator or negative exposure control remains open.

Our goal is to present and justify a regression-based method using an exposure-based indicator that can be used to reduce residual confounding in observational studies. Our main result is that the effect estimator from the "extended" model (with the indicator) tends to be less biased than that from the "final" model (without the indicator). That is, adding an exposure-based indicator to the model is expected to reduce residual confounding, given our assumptions. An important assumption is that measurement error and confounding are relatively unimportant. If this assumption is uncertain, the method can be used as the basis for sensitivity analyses to partially correct for bias.  Our approach differs from others because it uses an exposure-based rather than an outcome-based control variable in regression analyses.

**Methods**

We state our assumptions and then show that the effect estimator should be less biased when the future indicator is included in the model than when not included.

*Assumptions*:

We use directed acyclic graphs (DAGs) to summarize assumed causal relationships. Many syntheses of the terminology, construction and use of DAGs are available, so we do not repeat them here (4, 14-17). As in previous work (8), we illustrate using a time-series study with ambient air pollution on day $t$ as the exposure of interest ($X_t$) and emergency department visits on day $t$ as the health event ($Y_t$), although results apply more generally. We let $C_t$ represent controlled confounders (measured on day $t$ or before). The concern is that an unmeasured, perhaps unrecognized, confounder or covariate is associated with $X_t$ and is a cause of $Y_t$ (e.g., $U_t$ in Figure 1). We assume that any such covariate is associated with the exposure both on day $t$ ($X_t$) and the day after the health event ($X_{t+1}$). Throughout, we assume that measurement error and misspecification of functional form are negligible, so any model "misspecification" stems primarily from omitting one or more confounders ($U_t$, possibly vector-valued).

We base derivation of our results on two key assumptions (A1 − A2) about the pattern of causal effects, consistent with those summarized in Figure 1.

A1)  $Y_t(x_t, c_t, u_t) = \beta_o + \beta_1 x_t + \beta_2 c_t + \beta_3 u_t + \varepsilon_t$, where $\mathrm{E}[\varepsilon_t| X_t=x_t, C_t=c_t, U_t=u_t] = 0$;

A2)  The DAG in Figure 1 is correct; in particular, $X_{t+1}$ is independent of $Y_t$, conditional on $X_t$, $C_t$ and $U_t$, but like $X_t$, $X_{t+1}$ is associated with $\mathrm{U}_t$ (basic property).

The expression in A1 is interpretable as a structural equation, where $Y_t(x_t, c_t, u_t)$ is the counterfactual value of $Y_t$, if $X_t$ were set to $x_t$, $C_t$ to $c_t$, and $U_t$ to $u_t$. If $U_t$ were measured, we could fit the regression model $Y_t = b_o + b_1 x_t + b_2 c_t + b_3 u_t + \varepsilon_t$, to consistently estimate $\beta_1$, the effect of $X_t$ on $Y_t$, the analytic objective.

Our other assumptions are:

A3)  Model misspecification is due primarily to omission of the confounder $U_t$; other types of misspecification are negligible.

A4)  The joint distribution of ($\varepsilon$, $X_t$, $C_t$, $X_{t+1}$, $U_t$) is a stationary, ergodic process; and, regularity

conditions(18-20) hold.

A5)  $E[U_t / X_t=x_t, C_t=c_t, X_{t+1}=x_{t+1}] = \alpha_1 x_t + \alpha_2 c_t + \alpha_3 x_{t+1}$ .

A6)  $E[X_{t+1}/ X_t=x_t, C_t=c_t] = \gamma_1 x_t + \gamma_5 c$.

It is possible (but beyond the scope of this manuscript) to show that assumptions A5 and A6 are unnecessary so that $\alpha_1 x_t + \alpha_2 c_t + \alpha_3 x_{t+1}$ and $\gamma_0 + \gamma_1 x_t + \gamma_5 c$, are merely certain orthogonal projections, which exist quite generally even if the expectations have substantially more complicated form.

Our objective is now to derive expressions for bias when the confounder $U_t$ is omitted from the regression model.

*Mean of $Y_t$ conditional on $X_t$, $C_t$, and $X_{t+1}$, or on $X_t$ and $C_t$.*

To derive formulas for the magnitude of residual confounding due to omission of one or more covariates

($U_t$), we first consider the mean of $Y_t$ conditional on $X_t$, $C_t$, and $X_{t+1}$ or on $X_t$ and $C_t$.

To find an expression for $E[Y_t/X_t=x_t, C_t=c_t, X_{t+1}=x_{t+1}]$, we first note that, in general,

$E[Y_t/X_t=x_t, C_t=c_t, X_{t+1}=x_{t+1}] = E_{U_t|X_t, C_t X_{t+1}}[E[Y_t/X_t=x_t, C_t=c_t, U_t=u_t, X_{t+1}=x_{t+1}]]$.  Furthermore,

$E[Y_t/X_t=x_t, C_t=c_t, U_t=u_t, X_{t+1}=x_{t+1}]$ does not depend on $X_{t+1}$, given $X_t$, $C_t$, and $U_t$ by (A1) so we can simplify

by dropping $X_{t+1}$ and write:

$$E[Y_t/X_t=x_t, C_t=c_t, X_{t+1}=x_{t+1}] = E_{U_t|X_t, C_t X_{t+1}}[E[Y_t/X_t=x_t, C_t=c_t, U_t=u_t]]$$

$$= E_{U_t|X_t, C_t X_{t+1}}[\beta_o + \beta_1 x_t + \beta_2 c_t + \beta_3 u_t]$$

$$= \beta_o + \beta_1 x_t + \beta_2 c_t + \beta_3 E_{U_t|X_t, C_t X_{t+1}}[u_t]$$

$$= \beta_o + \beta_1 x_t + \beta_2 c_t + \beta_3(\alpha_1 x_t + \alpha_2 c_t + \alpha_3 x_{t+1})$$

$$= \beta_o + (\beta_1 + \beta_3 \alpha_1)x_t + (\beta_2 + \beta_3 \alpha_2)c_t + \beta_3 \alpha_3 x_{t+1} \qquad (1)$$

where the fourth equality follows by substitution of (A5). A similar evaluation gives:

$$E[Y_t/X_t=x_t, C_t=c] = \beta_o + (\beta_1 + \beta_3 \alpha_1)x_t + (\beta_2 + \beta_3 \alpha_2)c_t + \beta_3 \alpha_3 E_{U_t|X_t, C_t X_{t+1},}[x_{t+1}]$$

$$= \beta_o + \beta_3\alpha_o + (\beta_1 + \beta_3\alpha_1 + \beta_3\alpha_3\gamma_1)x_t + (\beta_2 + \beta_3\alpha_2 + \beta_3\alpha_3\gamma_2)c_t \qquad (2)$$

*Magnitude of confounding*:

If we fit the "extended" model, $E[Y_t/X_t=x_t, C_t=c_t, X_{t+1}= x_{t+1}] = b_o + b_1x_t + b_2c_t + b_3x_{t+1}$, say using least squares, the method of maximum likelihood, or generalized method of moments then

$$\widehat{b_1} \to \beta_1 + \beta_3\alpha_1 \qquad (3)$$

where $\to$ means convergence in probability. This follows from Equation 2 and usual linear regression results (e.g., under assumptions A1-A6). Similarly, if we fit the "final" model, $E[Y_t/X_t=x_t, C_t=c] = d_o + d_1x_t + d_2c_t$, then:

$$\widehat{d_1} \to \beta_1 + \beta_3\alpha_1 + \beta_3\alpha_3\gamma_1. \qquad (4)$$

These results show that $\widehat{b_1}$ and $\widehat{d_1}$ are not consistent estimators of parameter $\beta_1$, but have biases:

$B_1 = \beta_3\alpha_1$ for the extended model and $\qquad (5)$

$B_2 = \beta_3\alpha_1 + \beta_3\alpha_3\gamma_1$ for the final model. $\qquad (6)$

Here, "bias" refers to the difference between $\beta_1$ and the large sample limit of its estimator. Compared with the bias $B_1$ for the extended model, $B_2$ includes the additional term $\beta_3\alpha_3\gamma_1$. Thus, bias for the final model will be in the same direction but larger than that for the extended model, provided: the change of $X_t$ is in the same direction, on average, as the change of $X_{t+1}$ for an increase in $U_t$ (measured by signs of $\alpha_1$ and $\alpha_3$), and that $X_t$ is positively associated with $X_{t+1}$ ($\gamma_1 > 0$). When these two conditions are plausible, then we expect the magnitude of confounding in the final model to exceed that in the extended model by $\beta_3\alpha_3\gamma_1$.

*Confounding correction – sensitivity analyses*

The arguments above imply that $(\widehat{d_1} - \widehat{b_1})$ obtained by fitting Equations (3,4) will converge in probability to $\beta_3\alpha_3\gamma_1$, so if $\lambda^*$: is the value of $\lambda$ such that $\alpha_1 = \lambda^*\alpha_3\gamma_1$ then $\widehat{b_1} - \lambda^*(\widehat{d_1} - \widehat{b_1}) = \widehat{b_1}(1 + \lambda^*) - \lambda^*\widehat{d_1}$ will converge in probability to the parameter $\beta_1$, which is the causal effect of $x_1$ under assumption A1. Our confounding-corrected estimator is:

$$\widehat{\beta_{1,c}} = \widehat{b_1}(1 + \lambda^*) - \lambda^*\widehat{d_1} = \widehat{b_1} - \lambda^*(\widehat{d_1} - \widehat{b_1}) \qquad (7)$$

Even though $\gamma_1$ can be estimated as the linear association of $x_t$ with $x_{t+1}$, $\alpha_1$ and $\alpha_3$ are not identified and $\lambda^*$ will generally be unknown. However, $\lambda$ reflects the association of the conditional mean of $U_t$ with $X_t$ relative to its association with $X_{t+1}$. Absent adjustment for other covariates, $\lambda^*$ has a particularly simple interpretation, using $\rho$ – the (estimable) correlation coefficient of $X_t$ with $X_{t+1}$. If $\lambda^*$ if greater than $1/\rho$, then the mean of $U_t$ has a stronger linear association with $X_t$ than with $X_{t+1}$; if $\lambda^*$ equals $1/\rho$ then $U_t$ is equally associated with $X_t$ and $X_{t+1}$; if $\lambda^*$ is less than $1/\rho$ $U_t$, has a weaker association with $X_t$ than with $X_{t+1}$; and, if $\lambda^* = 0$, $U_t$ is linearly associated with $X_{t+1}$ but has no linear association with $X_t$ (no confounding).

Values of $\lambda$ can thus be specified as part of a sensitivity analysis. One approach for sensitivity analyses, might be to start with $\lambda$ equal to $1/\rho$ and consider different values of $\lambda$ over a reasonably wide range.  If $\lambda=0$ is included, that would include the possibility of no confounding and negative values include the possibility of confounding in the opposite direction. In some cases, it may be reasonable to assume that $\alpha_1 \geq \alpha_3$ in which case use of $(1+\frac{1}{\rho})\widehat{b_1} - \frac{1}{\rho}\widehat{d_1}$ to estimate $\beta_1$ would likely under-correct for confounding, at least absent control of other covariates. The variance of $\widehat{b_1}(1+\lambda^*) - \lambda^*\widehat{d_1}$ is estimated by $(1+\lambda^*)^2\text{Var}(\widehat{b_1}) + (\lambda^*)^2\text{Var}(\widehat{d_1}) - 2\lambda^*(1+\lambda^*)\text{Cov}(\widehat{b_1},\widehat{d_1})$, if $\lambda^*$ is treated as constant.

*Alternative Assumptions*

These bias equations also hold in the log-linear case, assuming that $Y_t$ is Poisson, and that $U_t$ has a Gaussian distribution (Appendix). It is possible to show that one can use assumptions weaker than (A5) and (A6) by using orthogonal projections and results concerning convergence of generalized method of moments estimators (18-20). However, we do not pursue that generality here.

*Simulations*

To assess the performance of our confounding-corrected estimator in finite samples and in log-linear models that are often used for health events, we conduct a series of simulations. (We also evaluated linear models, and found that the confounding-corrected estimator was less biased than the uncorrected estimator; however, we emphasize log-linear models here as they are commonly used in environmental epidemiology.) We assess the ability of this approach to partially correct for residual confounding using data from ongoing time-series studies of air pollution and daily emergency department visits (EDV) analyzed using Poisson regression. We use simulations so that the true causal relationships will be known. To make the simulations realistic, the "true" expected counts are the model-predicted counts of daily EDV for asthma (the health event) obtained by fitting a Poisson model to real, observed data for a recent 10 year period (1995-2004) in Atlanta. We use observed, 8-hour maximum ozone levels lagged 1 day as the air pollutant of interest (Table 1). To reduce heterogeneity, we restrict analyses to the warm season (May–October).

Analyses use the following log-linear Poisson model:

$$\log(E(Y_t)) = \beta_0 + \beta_1 x_t + \beta_2 c_t \qquad\qquad (8)$$

where $\beta_o$, $\beta_1$ and $\beta_2$ (a vector) are parameters and $c_t$ is a vector of controlled covariates including: linear, quadratic and cubic terms for time $t$ (day numbered from 1 to 185 for each 6 month period); linear, quadratic and cubic terms for the moving average of minimum temperature lagged 1day; indicators for temperature category ($1^{o}C$) lagged 1 day; indicators for day-of-week; indicators for month; and year; and product terms between the year and time terms. EDV counts ($Y_t$) are assumed to be Poisson with mean given by Equation 1. This model is similar to previously-used models (8, 21).

We fit this Poisson model using Proc Genmod in SAS (Version 9.3) to the observed counts to obtain model-predicted daily counts which we treat as the truth. This estimation approach is justified because the generalized method of moments includes maximum likelihood estimators if the moment functions and the score equations (22). For simulations with an assumed non-null air pollution effect (Scenarios 1B-6B), the coefficient ($\beta_1$) of $X_t$ was 0.0257 to calculate the model-predicted counts; for simulations with no assumed air pollution effect (Scenarios 1A-6A), we set $\beta_1 = 0$. We next generate simulated daily counts of EDV using a Poisson distribution with mean equal to model-predicted counts. We then analyze each simulated data set using models that include $X_t$ but not $X_{t+1}$, *and* models that include both $X_t$ and $X_{t+1}$.

Analyses are then conducted that miss-specify the analytic model by intentionally omitting one or more covariates (scenarios 2-6).  By omitting *known* covariates, we simulate bias that would occur if a covariate with a realistic distribution and realistic associations with exposure and outcome had been inadvertently omitted. Scenarios 2-6 differ because each omits a different covariate from the full, correct model (Tables 3-4, second column).  In our simulations $\widehat{\beta_1}$ refers to the estimator from the model *without* the future indicator and $\widetilde{\beta_1}$ is from the model *with* it. We calculate the bias in our estimators as the (median) log rate ratio $(\widehat{\beta_1})$ estimated with the misspecified model (e.g., a covariate omitted) minus the true $\beta$ ($X_t$-coefficient in the model used to generate the simulated data).  Finally, we compare the two estimators $\widehat{\beta_1}$ and $\widetilde{\beta_1}$ using: the median bias and the mean squared error.

**Results**

As shown in Table 3, a small to moderate bias in the log rate ratio was introduced in scenarios 2A - 6A by dropping: day-of-week; time; maximum temperature; both time and month, maximum temperature, and both minimum and maximum temperature variables, respectively (column 3 of Table 3).  In scenarios 3A-6A, the bias and mean squared error were slightly to moderately reduced when the Future Indicator was included in the regression model.

In scenario 2A, the bias was perhaps slightly worse when the future indicator was included; however, this is expected since most of the omitted day-of-week indicators (6 out of 7) were associated in opposite directions with the future indicator compared with the exposure, after adjustment for the other model covariates. We found similar results under the non-null (Table 4).

Thus, both under the null and non-null inclusion of the ozone level one day after the health event tended to at least somewhat reduce the bias due to confounding, when the exposure and future indicator were similarly associated with the omitted confounder.

**Discussion**:

We have shown that in log-linear models estimators in the extended model are expected to be less biased than those in the corresponding "final" model without the indicator, in the presence of unmeasured (omitted) confounders like those we considered.  Our simulations demonstrate this result empirically for log-linear models using actual covariates that we intentionally omitted, suggesting that one can often, but not invariably, expect use of the extended model to produce less biased estimates than the naive model, under often plausible assumptions. We also found a similar reduction of bias in the linear regression context (results not shown), but emphasize the log-linear case since that approach is common in practice.

Furthermore, we have suggested a simple, easily interpretable approach to sensitivity analyses.  Our approach, although somewhat like that of Lipsitch et. al (11), applies to the situation when known confounders are controlled by use of a regression model whereas theirs uses stratification. It also has similarities with the correction method of Tchetgen (13), Schuemie et al (23) and Richardson et al (24), but our approach uses an exposure-based indicator/control rather than an outcome control. This difference is important, as assumptions for an approach using an exposure-based indicator may hold whereas those for an outcome-based negative control could fail, or conversely.  Our approach should also work using negative-exposure controls as described by Liptsitch et al (11) which may differ technically from the future indicator; we emphasized the future indicator since temporality considerations help make needed assumptions plausible in  environmental time-series studies. Our approach is based on the parameter ($\lambda$) which is interpretable as a measure of the association of $U_t$ with $X_t$ compared to that with $X_{t+1}$ relative to the correlation ($\rho$) between $X_t$ and $X_{t+1}$. Values of $\lambda$ exceeding $1/\rho$ are consistent with $U_t$ having a stronger linear association with $X_t$ than with $X_{t+1}$ and values less than $1/\rho$ are consistent with a weaker association.

Although we found one scenario under the null and one under the non-null (2A or 2B, where inclusion of the future indicator did not lead to reduced bias, these two scenarios are consistent with expectations. In particular, a key assumption in our proof that the extended model should yield less biased estimators was that the uncontrolled (omitted) confounder be positively (or negatively) associated with *both* $X_t$ and $X_{t+1}$. However, the day-of-week indicators (Scenarios 2A and 2B) are perhaps somewhat unusual in that the association, for most days, is in the opposite directions for $X_t$ and $X_{t+1}$. For example, Sundays are associated with lower ozone levels on that day ($X_t$), but with higher levels the next day ($X_{t+1}$) – a Monday when traffic is back to weekday levels. Theoretically we should then expect no improvement by including the future indicator if confounding is due to omission of variables like day-of-week.  Thus, consideration of substantive issues and prior knowledge is important (25).

We derived results derived for the future indicator $X_{t+1}$, but the derivation mainly relied on the assumptions that $X_{t+1}$ was associated with $U_t$ (in the same direction as $X_t$) and was independent of $Y_t$ given $U_t$ and $X_{t+1}$.  Thus, the approach can be extended to justify the usual approach to correct for confounding given a surrogate variable for a confounder $U_t$. In other words, an extension can provide an alternative justification of the common practice of controlling for a surrogate of a confounder when the confounder itself is not measured directly, and suggests a formula for further correcting for confounding in sensitivity analyses.

Since the degree to which $U_t$ is associated with $X_t$ and $X_{t+1}$ is generally unknown ($\lambda$ not identified), the sensitivity analysis can be an important component of the approach.  We note that if no correction is used and no sensitivity analyses are presented or other allowance made for residual confounding, then the assumption is that there is no residual confounding – a rather strong assumption.

The approach we have described emphasizes confounding and is based on the assumption that confounding is the predominant source of bias.  However, if exposure measurement error is an important source of bias, then the future indicator may, in fact, be correlated with the poorly measured true exposure and inclusion of the future indicator could worsen the bias. Again, consideration of substantive issues and a prior knowledge is important.

In summary, we have presented a method to reduce residual confounding of effect estimators, even if the uncontrolled confounders are unmeasured. The approach assumes availability of an indicator variable with two key properties and that the unmeasured confounders have the same direction of association with both the actual exposure and the indicator.  We have argued that, for environmental studies such as those of air pollution and health effects, the air pollutant level after the health effect has occurred may be have the properties needed for such an indicator. Although not the primary focus, we have also suggested a method for sensitivity analyses that extends that of Lipsitch et al (11) since it can be used when the known confounders are controlled by regression models.

Table 1:   Description of Observed Data

| Variable | Mean (SD) | Median/ Min/  Max |
|---|---|---|
| Daily Asthma ED visits | 50.2  (21.2) | 46   /   6/   144 |
| Daily 8-hour, O3[1] | 2.28  (0.93) | 2.22/ 0.11 /  5.56 |
| Daily Max Temp (C) | 28.4  (4.44) | 29   /   11 /   39 |
| Daily Min Temp ( F) | 18.2  (4.35) | 19   /    1/   26 |

[1]O3 is measured in units of 25 ppb, *roughly* equal to its standard deviation.

Table 2:  Symbols and Notation

| Symbol | Represents |
|---|---|
| $Y_t$, | Random variable - the dependent variable at time $t$ |
| $X_t$ ,$x_t$ | Random variable – the exposure of primary interest at time $t$; $x_t$ denotes the value of $X_t$ |
| $C_t$ ,$c_t$ | Random variable – the covariates  at time $t$;  $c_t$ denotes the value of $C_t$ |
| $X_{t+1}$,$x_{t+1}$ | Random variable – the Future indicator at time $t$ ;  $x_{t+1}$ denotes the value of $X_{t+1}$ |
| $U_t$,$u_t$ | Random variable – confounder (unmeasured) at time $t$ ;  $u_{t+1}$ denotes the value of $U_{t+1}$ |
| $\beta_o$, $\beta_1$, $\beta_2$, $\beta_3$ | Parameters in model: E[$Y_t$/ $X_t$=x,$C_t$=c,$U_t$=u] = $\beta_o$+ $\beta_1 x$+ $\beta_2 c$+ $\beta_3 u$; note - $\beta_2$ is a row-vector of parameters, and $C_t$ a column-vector of covariates. |
| $\alpha_o$, $\alpha_1$, $\alpha_2$, $\alpha_3$ | Parameters in model: E[$U_t$/ $X_t$=x, $C_t$=c,$X_{t+1}$=$x_t$] = $\alpha_o$+ $\alpha_1 x$+ $\alpha_2 c$+ $g_U(x_t, c_t, x_{t+1})$ |
| $\gamma_o$, $\gamma_1$, $\gamma_2$ | Parameters in model:  E[$X_{t+1}$/ $X_t$=$x_t$, $C_t$=$c_t$] =  $\gamma_o$+ $\gamma_1 x_t$+ $\gamma_2 c_t$+ $g_X(x_t, c_t, x_t)$ |
| b$_0$,b$_1$,b$_2$,b$_3$ | Parameters in miss-specified model: E[$Y_t$/ $X_t$=$x_t$,$C_t$=$c_t$,$X_{t+1}$=$x_{t+1}$] =  b$_o$+ b$_1 x_t$+ b$_2 c_t$+ b$_3 x_{t+1}$ that omits $U_t$. Note: the "extended" model includes $x_{t+1}$, the :"final" model does not. |
| $\widehat{b_1}$ | estimator of $b_1$ from the final model (*without* the future indicator) |
| $\widehat{b_1^*}$ | estimator of $b_1$ from the extended model (*with* the future indicator) |

Table 3: Simulation Results under Null Hypothesis, true effect of Air Pollutant $b_1 = 0$. (rev5)

| Scenario and Description | | Uncorrected Estimator | | Corrected Estimator | |
|---|---|---|---|---|---|
| Scenario | Type Analytic Error/ Misspecification | Bias[1]: Median $\widehat{b_1}$ - True $\hat{\beta}$ (SE ($\widehat{b_1}$)) | MSE | Bias[1]: Median $\widehat{b_1^*}$ - True $\hat{\beta}$ (SE ($\widehat{b_1^*}$)) | MSE |
| 1A | None | -0.0003    (0.0053) | 0.0053 | -0.0004    (0.0053) | 0.0053 |
| 2A | Omit day of week indicators | 0.0062   (0.0069) | 0.0082 | 0.0073    (0.0069) | 0.0093 |
| 3A | Omit continuous time variables $(t, t^2, t^3)$ | -0.0040  (0.0057) | 0.0067 | -0.0035   (0.0059) | 0.0064 |
| 4A | Omit continuous time variables *and* Month indicators | -0.0461  (0.0093) | 0.0465 | -0.0372  (0.0093) | 0.0375 |
| 5A | Omit Maximum Temperature Indicators | 0.0088  (0.0049) | 0.0100 | 0.0080   (0.0050) | 0.0094 |
| 6A | Omit Minimum *and* Max Temperature Variables | 0.0019  (0.0055) | 0.0058 | 0.0015   (0.0055) | 0.0057 |

Table 4 Simulation Results under Alternative Hypothesis, With True Non-null effect of Air Pollutant $b_1 = 0.0257$.

| Scenario and Description | | Uncorrected Estimator | | Corrected Estimator | |
|---|---|---|---|---|---|
| Scenario | Type Analytic Error - Misspecification | Bias[1]: Median $\widehat{b_1}$ - True $\hat{\beta}$ (SE ($\widehat{b_1}$)) | MSE | Bias[1]: Median $\widehat{b_1^*}$ - True $\hat{\beta}$ (SE ($\widehat{b_1^*}$)) | MSE |
| 1B | None | -0.0003    (0.0053) | 0.0053 | -0.0004    (0.0053) | 0.0053 |
| 2B | Omit day of week indicators | 0.0062   (0.0069) | 0.0082 | 0.0073    (0.0069) | 0.0093 |
| 3B | Omit continuous time variables $(t, t^2, t^3)$ | -0.0042  (0.0057) | 0.0078 | -0.0037  (0.0058) | 0.0075 |
| 4B | Omit continuous time variables *and* Month indicators | -0.0460  (0.0092) | 0.0068 | -0.0370  (0.0092) | 0.0065 |
| 5B | Omit Maximum Temperature Indicators | 0.0087   (0.0047) | 0.0097 | 0.0076  (0.0049) | 0.0090 |
| 6B | Omit Minimum *and* Max Temperature Variables | 0.0015  (0.0058) | 0.0055 | 0.0010   (0.0053) | 0.0055 |

Figure 1

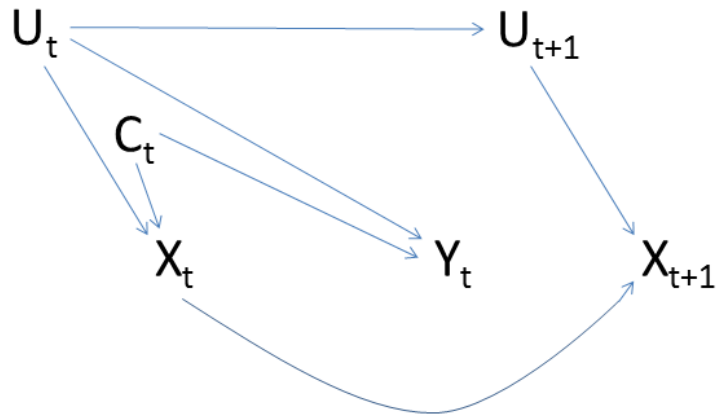

**Legend**: A DAG Consistent with Assumed Causal Relationships.  Figure 1 depicts the outcome ($Y_t$), exposure ($X_t$), modeled confounder ($C_t$), omitted confounder ($U_t$) at time t, and the exposure after the outcome has occurred ($X_{t+1}$; the future indicator).

**Appendix  – the log-linear case.**

We can derive results similar to those in the main text, based on alternative assumptions:

A1*)   $Y_t$ has mean $\exp(\beta_o + \beta_1 X_t + \beta_2 C_t + \beta_3 U_t)$, conditional on $X_t=x, C_t=c, U_t=u$.

This assumption is commonly used in modeling health outcomes measured as counts where an additional assumption is often that $Y_t$ is Poisson.

A2*)  the distribution of $U_t$ given $X_t=x$, $C_t=c_t$ and $X_{t+1}=x_{t+1}$ denoted by

$f_U(U_t|\beta, X_t=x_t, C_t=c_t, X_{t+1}=x_{t+1})$: has moment generating function $\Psi_U(\beta,x,c,f)$.

_Claim1B_. Assumptions (A1*) and (A2*) imply:

$E[Y_t/X_t=x_t, C_t=c_t, X_{t+1}=x_{t+1}]=\exp(\beta_o+\beta_1 x_t+\beta_2 c_t+log(\Psi_U(\beta_3;x_t,c_t,x_{t+1})))$,          (10A)

where $log(\Psi_U(\beta_3; x_t,c_t,x_{t+1}))$ is the moment generating function of the conditional distribution of $U_t$ given

$X_t=x_t, C_t=c_t, X_{t+1}=x_{t+1}$.

Proof of Claim1B:   As in the proof of Claim 1 (Appendix 1), we evaluate the expectation of $Y_t$ by integrating the joint distribution of $Y_t$ and $U_t$ given $X_t=x, C_t=c$, $X_{t+1}=x_{t+1}$ over $U_t$ to obtain:

$E[Y_t/X_t=x, C_t=c, X_t=x_{t+1}] = \iint y\ f_Y(y/X_t=x, C_t=c, U_t=u) f_U(u/X_t=x_t, C_t=c_t, X_{t+1}=x_{t+1})$ dyd$u$

$= \int\exp(\beta_o + \beta_1 x + \beta_2 c + \beta_3 u) f_U(u/ X_t=x_t, C_t=c_t, X_{t+1}=x_{t+1})$ d$u$          (11A)

The second equality follows from assumption (A1*) by carrying out the integration with respect to $y$ and where $f_U(u/ X_t=x, C_t=c, X_{t+1}=x_{t+1})$ is the conditional pdf of $U_t$. Rearrangement now shows that the moment generating function of $f_U(u/ X_t=x_t, C_t=c_t, X_{t+1}=x_{t+1})$ appears as a factor on the right hand side:

$E[Y_t/ X_t=x_t, C_t=c_t, X_{t+1}=x_{t+1}] = \exp(\beta_o + \beta_1 x + \beta_2 c) \int \exp(\beta_3 u) f_U(u/ X_t=x_t, C_t=c_t, X_{t+1}=x_{t+1})$d$u$          (12A)

Substituting $\Psi_U(\beta_3;x,c,f)$ for its defining integral gives:

$E[Y_t/ X_t=x_t, C_t=c_t, X_{t+1}=x_{t+1}] = \exp(\beta_o + \beta_1 x_t + \beta_2 c_t) \Psi_U(\beta_3;x_t,c_t,x_{t+1})$

$$= \exp(\beta_o + \beta_1 x + \beta_2 c + \log(\Psi_U(\beta_3; x_t, c_t, x_{t+1}))) \qquad (13A)$$

proving Claim 1B.

*Corollary 1*: If $U_t = \alpha_0 + \alpha_1 x_t + \alpha_2 c_t + \alpha_3 x_{t+1} + \varepsilon_{U,t}$ with $\varepsilon_{U,t}$ Gaussian, then $E[Y_t / X_t = x_t, C_t = c_t,$

$X_{t+1} = x_{t+1}] = \exp(\beta_0^* + (\beta_1 + \beta_3\alpha_1)x_t + (\beta_2 + \beta_3\alpha_2)c_t + (\alpha_3 + \beta_3\alpha_3)x_{t+1})$

<u>Proof</u>: The result follows by substitution, since $\Psi_U(\beta_3; x_t, c_t, x_{t+1}) = e^{\beta_3(\alpha_0 + \alpha_1 x_t + \alpha_2 c_t + \alpha_3 x_{t+1}) + \frac{1}{2}\sigma^2\beta_3^2}$.

<u>*Claim 1C*</u>. Assumptions (A1\*) and (A2\*), and those for Claim 1B and Corollary 1 imply:

$$E[Y_t / X_t = x_t, C_t = c_t] = \exp(\beta_0^* + (\beta_1 + \beta_3\alpha_1)x_t + (\beta_2 + \beta_3\alpha_2)c_t + log(\Psi_X((\alpha_3 + \beta_3\alpha_3); x_t, c_t)) \qquad (14A)$$

where $\Psi_X((\alpha_3 + \beta_3\alpha_3); x_t, c_t, x_{t+1})$ is the moment generating function of the conditional distribution of $X_{t+1}$ given $X_t = x_t, C_t = c_t$ with variable $(\alpha_3 + \beta_3\alpha_3)$.

<u>Proof of Claim1C</u>: Follows from Claim 1B and Corollary 1, by a proof similar to that used for Claim 1B.

*Corollary 2*: If, in addition to the assumptions of Corollary 1, $X_{t+1}$ is Gaussian with mean

$E[X_{t+1} / X_t = x_t, C_t = c_t] = \gamma_0 + \gamma_1 x_t + \gamma_2 c_t$ then $E[Y_t / X_t = x_t, C_t = c_t] = \exp(\beta_0^{**} + (\beta_1 + \beta_3\alpha_1 + \beta_3\alpha_3\gamma_1)x_t + (\beta_2 + \beta_3\alpha_2 + \beta_3\alpha_3\gamma_2)c_t)$

<u>Proof</u>: The result follows from Claim 1C by substitution of the moment generating function of $X_{t+1}$.

<u>*Note*</u>: Under the assumptions of Corollaries 1 and 2, the mean of Y is log-linear, in the variables $X_t, C_t,$ $X_{t+1}$ or $X_t, C_t$. Thus, if Y is Poisson, the maximum likelihood estimators of $\beta_1$ will converge as indicated in Equations 3 and 4 of the main text. Thus, the bias specified in Equations 5 and 6, will hold under the alternative assumptions A1\*, A2\*, Corollary 2 or 3, and A3-A6.